# ODESSA SCIENTIFIC SCHOOL OF RESEARCHERS OF VARIABLE STARS: FROM V.P.TSESEVICH (1907-1983) TO OUR DAYS

*Andronov I.L.*

Department of Mathematics, Physics and Astronomy, Odessa National Maritime University
tt_ari@ukr.net

ABSTRACT. The biography of Vladimir Platonovich Tsesevich (11.11.1907 – 28.10.1983), a leader of the astronomy in Odessa from 1944 to 1983, is briefly reviewed, as well as the directions of study, mainly the highlights of the research of variable stars carried out by the members of the scientific school founded by him. The directions of these studies cover a very wide range of variability types – "magnetic" and "non-magnetic" cataclysmic variables, symbiotic, X-Ray and other interacting binaries, classical eclipsers and "extreme direct impactors", pulsating variables from DSct and RR through C and RV to SR and M. Improved algorithms and programs have been elaborated for statistically optimal phenomenological and physical modeling.

Initially these studies in Odessa were inspired by ("with a capital letter") Vladimir Platonovich Tsesevich. who was a meticulous Scientist and brilliant Educator, thorough Author and the intelligibly explaining Popularizer, persevering Organizer and cheerful Joker - a true Professor and Teacher. He was "the Poet of the Starry Heavens".

## 1. V.P. Tsesevich (11.10.1907 – 28.10.1983)

Vladimir Platonovich Tsesevich was born in Kiev on 11.10.1997. His father Tsesevich Platon Ivanovich was a famous opera singer (bass voice). His beautiful songs may be found in the Internet. His mother Kuznetsova Elisaveta Aleksandrovna was an opera actress, later a pedagogue. V.P. Tsesevich started education on the Leningrad State University at the age of almost 15, and chosen variable stars as the main direction of his studies.

The first paper based on his observations (but he was not a co-author) appeared also in 1922, but the first his own paper appeared next year (totally 42 during his studentship). The name was written as "W.Zessewitsch" in the papers published in the "Astronomische Nachrichten".

The supervisor of his PhD studies was G.A.Tikhov, a famous astronomer. After Leningrad, in 1933-1937, he was a Director of the Tadjik Astronomical Observatory (now Institute of Astrophysics). In 1937-1942, worked in Leningrad. Evacuated to Dushanbe, where worked as a professor, and finally moved to the Odessa State University in 1944.

Under his supervision, the initially small astronomical observatory in Odessa has become one of the leading astronomical organizations.

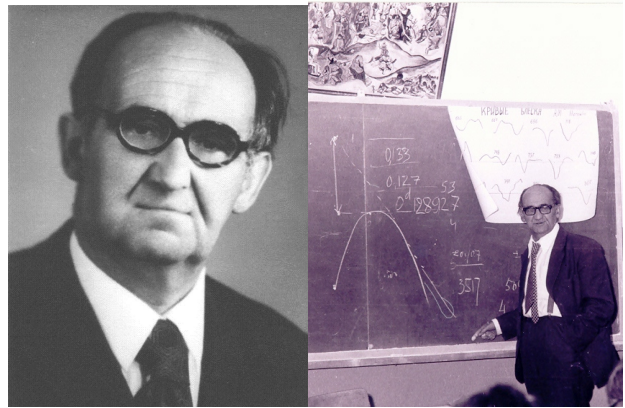

In 1948-50, his talent of an organizer was effectively realized in being a director of the Main Astronomical Observatory of the Ukrainian Academy of Sciences (1948-1950). In 1948, he was elected to be a corresponding member of the Ukrainian Academy of Sciences.

In our memory, remained a deep impression of his tremendous energy and enthusiasm of Vladimir Platonovich Tsesevich. He was not only an outstanding scientist and organizer of science, the founder of the scientific school of investigators of variable stars, but also a brilliant lecturer and popularizer of astronomy. The list of his 730 publications was compiled by Dziubina & Rikun (1988). However, in the ADS there is only a small part of them (142 for "Tsesevich", 31 "Tsessevich", 44 "Zessewitsch").

His lectures for students were bewitching and exciting. Among them, there were "Additional Chapters of Mathematical Physics" and two semesters of "Relativistic Astrophysics". He had read the lectures, as a poem, and it was exciting. He deduced numerous formulae without lecture notes, neglecting the bells, and consequently stopped, when difficult transition to some formula, sometimes lasting of few pages, came to the end. He taught not to be afraid of challenges and showed a process of creativity. Exacting to himself, he was exacting to others. He followed numerous new discoveries and ideas in astronomy, and proposed new directions of study. His last scientific passion was the magnetic cataclysmic binary system AM Herculis. The first papers showing its extreme exotic nature were published 40 years ago.

V.P.Tsesevich supervised more than 40 PhDs graduates, creating an effective scientific school and supported different scientific directions. V.P. Tsesevich was very different, somehow similar to the variable stars he studied.

Besides being a director of the Astronomical observatory of the I.I. Mechnikov Odessa State (currently National)

University (ONU) and the Chair of the Astronomical Department of the Physical Faculty of ONU, he lectured in some other institutes - of the Refrigerator Industry, in the Odessa Higher Engineering Marine School, in the Odessa Institute of Marine Fleet Engineers (currently Odessa State Maritime University (ONMU)). In the last institute, he lectured for 15 years and even was a Chair of the Department of Higher Mathematics (currently the Department of Mathematics, Physics and Astronomy).

Among these publications, there were numerous monographs, which may be subdivided into "printed catalogues" and "classical" monographs like "Variable Stars and Methods of Their Observations" (Tsesevich, 1970, 1980), famous collective monographs "Eclipsing variable Stars" (Tsesevich, 1971), "RR Lyrae-type Stars" (Tsesevich, 1969), both translated to English.

The most popular book is "What and How to Observe in the Sky" (Tsesevich, 1984). It has been issued six times, last time in 1984, and even now is one among the best books.

Not all readers became astronomers, but many people remember the book, and many students remember talented and emotional professional and public lectures. This had initiated passion to astronomy (and science generally) of many scientists, who now represent the base of our majestic science.

The asteroid 2498 (discovered on 23.08.1977 by N.S.Chernykh in the Crimean Astrophysical Observatory) was named "Tsesevich". "Named in honor of Vladimir Platonovich Tsesevich (1907–1983), former director of the Odessa University Observatory, renowned for his research on variable stars. He also studied the brightness variations of {433} Eros and is the author of a handbook for amateur astronomers".

The Google search for V.P.Tsesevich (in Russian) gives 3300 entries, from which we would like to point out some memories by Volyanska, Karetnikov & Mandel (2007), Vavilova (2017), Andronov (2003), Samus' (1988, 2007). The compilation of memories was prepared to his 100-th anniversary (Tsesevich, 2007).

**2. After V.P. Tsesevich**

During the last years of life of Vladimir Platonovich Tsesevich (11.11.1907-28.10.1983), the Astronomical observatory of the I.I.Mechnikov Odessa National (previously "State") University (ONU) became a widely known University center, which worked in a close connection with the Department of Astronomy.

Odessa was the only place in the country, where the director of the observatory and the chair of the Department of Astronomy was the same person (1871-2016).

At that time, there were few departments and smaller "sectors", which covered various directions of astronomy, so the students had a wide choice, and later worked in varies observatories and academic institutes.

Besides ONU, in different years, astronomers worked in other organizations in Odessa (N.I.Divari, O.E.Mandel in the Polytechnical Institute (now Odessa National Polytechnical University), V.A.Smirnov, L.V.Glazunova in the Odessa National Academy of Telecommunications, I.A.Klyus, V.V.Mikhalchuk in the Odessa State Maritime Academy, 5 astronomers in the Odessa National University; there is an Odessa Department of the Radioastronomical Institute of NASU). The astronomers from various organizations participate in events organized by the Odessa Astronomical Society (chair - M.I.Ryabov (56 papers)).

The secretary of the Department of Astronomy was V.G.Karetnikov (in 1983-2006 - the chair), who had a scientific working group at the observatory with a main direction on photometrical and spectrospoic study of classical eclipsing binary systems. The current number of publications listed in the ADS is 182. V.G.Karetnikov was a supervisor of the PhD Theses by E.V.Menchenkova (23 joint (only) papers/49 totally), V.V.Nazarenko (15/62), L.V.Glazunova (8/46), F.V. Sirotkin (4/27), G.V.Volkova-Manilova (2/13), S.M.Andrievsky (2/194), I.Kudzej (1/60), K.A.Antoniuk (0/102). Also in this group, worked S.V.Kutsenko (9), O.G.Lakinskaja (2). 16 papers were published with Yu.A.Medvedev. This group continued studies of eclipsing systems, started in Odessa by V.P.Tsesevich and A.M.Shulberg.

V.G.Karetnikov was the Chair of the Department of Astronomy (1983-2006), the Director of the Astronomical Observatory (1990-2006). During this period, the following Doctors of Science were graduated: N.S.Komarov (1990), I.S.Shestaka (1994), I.L.Andronov (1995), S.M.Andrievsky (2002), T.V.Mishenina (2005).

The traditionally largest department of the observatory on "Variable Stars" was supervised by Yu.S.Romanov (68 papers in the ADS) with collaborators working on analysis in Odessa and observing in Mayaki at the 7-th camera astrograph (V.P.Bezdenezhnyj (11), B.A.Murnikov (12), A.I.Pikhun (19), S.V.Kashuba). The majority of papers were devoted to photometrical, and spectral studies of the RR Lyr-type stars. In this Department, also actively worked in the Mayaki observational station, having own directions at 20" and AZT-3 telescopes, but with a monitoring th the 7-camera astrograph. Yu.S.Romanov was a supervisor of PhD theses of S.N.Udovichenko (13/48), A.V.Yushchenko (1/119), D.E.Mkrtichian (1/167), G.A.Garbuzov (2/34). Other active collaborators were Z.N.Fenina (16/41), L.P.Zaikova (5/9), A.N.Rudenko (3), O.P.Paramonova (2/11), L.E.Lysova (2/13), V.P.Murnikova (1/10), B.A.Murnikov (1/12) A.I.Movchan (1/16), A.S.Gadun (1/73) et al.

The first number is the number of joint paper in collaboration wit the head of the Department. A special program on studies of the Blazhko effect and determination of the moments of brightness maxima of RR Lyr-type stars was carried out by B.N.Firmanyuk (54) et al.

The second largest department "Astrospectroscopy" was supervised by N.S.Komarov (168 papers), where PhD theses defended T.V.Mishenina (159 papers), V.F.Gopka (83), T.N.Dorokhova (73), D.N.Doikov (11).

N.I.Dorokhov (71) had multiple trips to Dushak-Eregdag for photometrical observations of TT Ari, lambda Boo-type and other stars according to the international campaigns.

Later on, the largest Departments "Variable Stars" and "AstroSpectroscopy" were merged into one Department of "Physics of Stars and Galaxies", initially supervised by N.S.Komarov, and, since 2003, by T.V.Mishenina. The scientific secretary of AO is A.V.Dragunova (25). An important contribution made S.I.Belik (40), V.A.Pozigun (25), V.F.Karamysh (14), L.F.Orlova (14) et al.

In this Department, currently work all Doctors of Sciences of the Observatory - S.M.Andrievsky (194), V.V.Kovtyukh (172), T.V.Mishenina (159), A.I.Zhuk (147), S.A.Korotin (113). Traditionally partially worked the Professors of the Department of Astronomy - V.G.Karetnikov (182), I.L.Andronov (363, before 2006), E.A.Panko (64, after 2016). In recent decade, PhDs habilitated T.I.Gorbaneva (34), A.Chepizhko (7), L.L.Chinarova (88), V.A.Yushchenko (17), F.A.Chekhonadskikh (14), M. Eingorn (66), A. Chopovsky (11), A.L.Sukharev (21), A.I.Donskykh (11).

Not surprisingly, that astronomers – Doctors of Science posess top places at the scientific rating in the I.I.Mechnikov National University and give the highest contribution to the scientific outcome.

The third large Department from the times of V.P.Tsesevich is on "Space Research", initially supervised by Yu.A.Medvedev (50) and currently by N.I.Koshkin (50). Currently there work PhDs P.P.Sukhov (50), V.V. Troianskyi (13) the Associate Professor A.A.Bazey (5), and actively working L.S.Shakun (26), S.L.Strakhova (16), S.M.Melikyants (10), V.V.Dragomiretsky (10), E.A. Korobeinikova (5) and others. The major direction is on astrometry photometry of satellites and asteroids.

In the times of V.P.Tsesevich, there was a separate sector of astrometry, where worked M.Yu.Volyanskaya (21), M.I.Myalkovskiy (12), A.P. Chelombit'ko (6), V.V.Zhukov (16), O.S.Shakhrukhanov (3), N.V.Bazey (2). There was a sector of "Astronomical Instrument Making". N.N.Fashchevskiy (21) got his PhD in Astrophysics based on the observations obtained at the telescope he designed. E.A. Depenchuk (21), V.N.Ivanov (13), M.G.Arkhipov (8), L.S.Paulin (4), A.F.Pereverzentsev (5), A.V.Ryabov (21) made a great contribution in creating telescopes of the Odessa Observatory.

The supervisor of the special sector on "Meteors and comets" was professor E.N.Kramer (135), also there worked I.S.Shestaka (98), A.K.Markina (33), V.I. Musij (22), and later Yu.M.Gorbanev, who is the current leader of the working group.

It should be mentioned that these reminiscenses do not follow all the collaborators working in ONU or graduated in ONU, but working in other institutions.

I.L.Andronov (363) is the last PhD student (No.~41), who habilitated under the supervision of V.P.Tsesevich. He continued the direction of studies of variable stars - initially magnetic (and non-magnetic) cataclysmic, later X-Ray binaries, classical eclipsing binaries, rare "Extreme Direct Impactors", symbiotic, Mira-type, Semi-regular, RV, RR and other pulsating and newly discovered variables. The observations are carried out within a campaign "Inter-Longitude Astronomy" (ILA), which is based on temporarily working groups from different countries. Besides, an expert system for advanced modeling of the time series was elaborated, which is oriented onto general tasks, as well as on the specifics of different types of variability of stars. Generally 1900+ stars were studied in this group. The following eight PhD theses were habilitated: L.S.Kudashkina (1997, 55 papers), S.V.Kolesnikov (2000, 79), V.I.Marsakova (2000, 54), A.V.Halevin (2000, 36), A.V.Baklanov (2005, 65), V.V.Breus (2013, 25), L.L.Chinarova (2014, 88), M.G.Tkachenko (2017, 13). Currently, this group (4 in the Odessa National Maritime University, 3 in the Odessa National University; 2 effectively work in astronomy outside), together with S.N.Udovichenko (48 papers, supervisor Yu.S.Romanov), represent in the Ukrainian Universities the main direction of studies of V.P.Tsesevich - the photometric variability of stars of different types. There are prominent graduates from Odessa, now actively working outside Ukraine.

### 3. "Inter-Longitude Astronomy" (ILA) Campaign
### 3.1. The ILA Group

Continuing the scientific school founded by V.P. Tsesevich, the variable stars of different types are studied on the base of temporarily working groups in collaboration with astronomers from Korea, Germany, USA, Greece, France, Poland, Slovakia, Hungary, Spain, Portugal, Kazakhstan and other countries.

For the time series analysis, we make long-term photometric and polarimetric monitoring of the group of selected "key" objects as well as use the data from orbital and ground-based observatories.

In Ukraine, this group consists of 4 astronomers in the Odessa National Maritime University (Prof. Dr. Ivan L. Andronov, Dr. Vitalii V. Breus, Dr. Larysa S. Kudashkina, Dr. Mariia G. Tkachenko), 3 staff members in the Odessa National University (Dr. Lidiia L. Chinarova, Dr. Sergei V. Kolesnikov and Dr. Vladyslava I. Marsakova) and 2 students (Kateryna D. Andrych, Dmytro E. Tvardovskyi).

The total number of the stars studied in this group exceeds 1900, the number of articles of coauthors of this project, referred in the database "Astronomy Data System" (ADS), exceeds 400 (for the leader of group: 370 since 1980, 24 in 2015-2017). Members of the group from Odessa defended 9 PhD Theses and a 1 of Dr. Sci.

This project is called "Inter – Longitude Astronomy" (ILA), the funding to the participants is due to corresponding universities. The reviews of highlights of the project were published by Andronov et al. (2003, 2010, 2017). Also results are included in the national projects "Ukrainian Virtual observatory" (UkrVO) (Vavilova et al., 2011, 2012) and "AstroInformatics" (Vavilova et al., 2017).

### 3.2. "Polar" (Gravi-Magnetic Rotators)

Study of cataclysmic binary systems with magnetic white dwarfs (AM Her, QQ Vul, BY Cam, V1432 Aql, V808 Aur). This monitoring started in 1978 photometrically and in 1989 polarimetrically (Kolesnikov and Andronov, 2017). Every year new observations of magnetic cataclysmic variables with quickly rotating white dwarfs are carried out to study rotational evolution and check model of precession proposed earlier and also to study observational appearance of dependence of structure of the accretion stream on the angle between the magnetic axis and the line of centers. Systems for long-term monitoring are intermediate polars: BG CMi, MU Cam =1RXS J062518.2+733433, FO Aqr, AO Psc, 1RXS J063631.9 +353537, 1RXS J070407.9 +262501, 1RXS J180340.0 +401214, 1RXS J192626.8 +132153, 1RXS J213344.1 +510725, PQ Gem, V405 Aqr, EX Hya.

Recently, the spin-up of the magnetic white dwarf is discovered in the system MU Cam with characteristic with characteristic time-scale of $\tau = P/|dP/dt| = (170 \pm 1.5)$ thousand years. This value is 30 times less than 4.71 million years, which is observed in other intermediate polar EX Hya, but only twice less than 290 thousand years in BG CMi (Andronov et al., 2017).

### 3.3. "SuperHumper"

*Cataclysmic binary systems (nova-like and dwarf novae) with precessing accretion disks and QPOs.* Main target for the international monitoring: TT Ari, where we studied transitions between positive and negative superhumps and discovered a state of "excitation of quasi-periodic oscillations" (Kim et al., 2008). Other important targets are MV Lyr and V1084 Her. Very interesting are "Transient Periodic Oscillations" in DO Dra (Andronov et al., 2008).

### 3.4. "SSS" (Super-Soft Sources)

*Research of rapid variability* of the interacting binary systems - s*uper-soft sources* of X-ray radiation (V Sge, QR And).

### 3.5. "NewVar" (New Variable)

*Discovery and complete study of new variables* in the selected fields on the base of the new special observations. Initially, the studies of new variables were carried out on photographic plates of the Odessa, Moscow and Sonneberg plate collections. Then we studied 863 faint variable candidates from the Hipparcos-Tycho mission. Currently, we study newly discovered variables in the field of the main targets (e.g. Kim et al., 2004).

### 3.6. "Eclipser"

*Mathematical modeling of the asymmetric interactive binary systems.* The method NAV was elaborated, which allows phenomenological modeling of complete light curve. For detached MS systems, it allows to determine physical parameters as well (Andronov et al., 2015). The minima are fitted not only for EA-type systems, but also for EB and EW (Tkachenko et al., 2016). Especially interesting are "Extreme Direct Impactors" (Andronov and Richter, 1987, Andronov et al., 2010).

### 3.7. "Stellar Bell"

*Careful photometric researches and modeling* of the known variables, including pulsating variables with "combined", or "alternatively operating" types. The review paper with results of our researches of long-period variables was presented by Andronov et al. (2014).

### 3.8. "TSA" (Time Series Analysis)

*Elaboration and improvement of the algorithms and programs for the analysis of multicomponent signals.* The algorithms allow studies of irregularly spaced data, which are characteristic for photometrical surveys from space and ground-based observatories, and provide determination of phenomenological characteristics with much better accuracy (Andronov, 2005, Andronov, Tkachenko & Chinarova, 2016).

### 4. Other current variable star researches in Odessa

Astronomical Observatory of ONU has a great "Sky Patrol" plate collection, which is the third in the world according to the number of plates after Harvard (USA) and Sonneberg (Germany).

Hundreds of papers were published from 1956, the last one (in the epoch of visual estimates of brightness on plates) by Chinarova & Andronov (2000). As the best accuracy is ~0.08$^m$, this is a great resource for studies of historical light curves of large-amplitude (>0.5$^m$) stars.

Active CCD observations of DSct and RR pulsating variables (and stars in the field) are carried out at the 48-cm telescope AZT-3 by S.N.Udovichenko (2012, 2015). Among his 48 papers, 18 are based on the CCD observations since 1997.